\documentstyle[sprocl]{article}

\begin{document}

\title{TRANSLATIONAL AND ROTATIONAL PROPERTIES
OF ANTISYMMETRIC TENSOR FIELDS}

\author{VALERI V. DVOEGLAZOV}

\address{Escuela de F\'{\i}sica, Universidad Aut\'onoma de Zacatecas\\
Apartado Postal C-580, Zacatecas 98068 Zac., M\'exico\\
E-mail: valeri@ahobon.reduaz.mx\\
URL: http://ahobon.reduaz.mx/\~~valeri/valeri.htm}

\maketitle

\abstracts{Recently, several discussions on the possible observability
of 4-vector potential have been published in literature.
Furthermore, several authors recently claimed existence of
the helicity=0 electromagnetic field.
We re-examine the theory of antisymmetric tensor
field and 4-vector potentials. We study the massless limits too.
In fact, a theoretical motivation for this venture is old papers of
Ogievetski\u{\i} and Polubarinov, Hayashi, and Kalb and Ramond,
which are widely accepted by physics community.\linebreak
This paper is based on two poster presentations
``About the Longitudinal Nature of the Antisymmetric Tensor Field after
Quantization" and  ``Normalization and $m\rightarrow 0$ Limit of the Proca
Theory" at the Workshop ``Lorentz Group, CPT and Neutrinos" (Zacatecas,
M\'exico, June 23-26, 1999).}{}{}

\section{Introduction}

Since 1984 I was concerned with the puzzle of the so-called Kalb-Ramond
field~\cite{Ogievet,Hayashi} and the problem of higher spins. Later,
thanks to Dirac~\cite{Dirac} and Weinberg works~\cite{Weinberg} I learnt
that a number of theoretical aspects of electrodynamic theory deserve more
attention.  First of all, they are:  the problem of ``fictious photons of
helicity other than $\pm j$, as well as the indefinite metric that must
accompany them"; the renormalization idea, which ``would be sensible only
if it was applied with finite renormalization factors, not infinite ones
(one is not allowed to neglect [and to subtract] infinitely large
quantities)"; contradictions with the Weinberg theorem ``that no symmetric
tensor field of rank $j$ can be constructed from the creation and
annihilation operators of massless particles of spin $j$",\, {\it etc.}
Moreover, it appears now that we do not yet understand many specific
features of classical electromagnetism, first of all, the  problems of
longitudinal modes, of the gauge, of the Coulomb action-at-a-distance, and
of the Horwitz' additional invariant parameter,
refs.~\cite{Dvoegl,Ahlu,Staru,Chub,Horwitz}. In this presentation I
re-examine the theory of   4-potential field and antisymmetric
tensor field of the second rank. The concept of the
Ogievetski\u{\i}-Polubarinov-Kalb-Ramond field is simplified considerably.

\section{4-potentials and Antisymmetric Tensor Field}

The spin-0 and spin-1 particles can be constructed by taking the direct
product of the spin-1/2 field functions~\cite{BW}.
Let us firstly repeat the Bargmann-Wigner procedure of
obtaining the equations for bosons of spin 0 and 1.
The set of basic equations for  $j=0$ and $j=1$ are written, e.g.,
ref.~\cite{Lurie}
\begin{eqnarray} \left [
i\gamma^\mu \partial_\mu -m \right ]_{\alpha\beta} \Psi_{\beta\gamma} (x)
&=& 0\quad,\label{bw1}\\ \left [ i\gamma^\mu \partial_\mu -m \right
]_{\gamma\beta} \Psi_{\alpha\beta} (x) &=& 0\quad.\label{bw2}
\end{eqnarray}
We expand the $4\times 4$ matrix wave function into the antisymmetric and
symmetric parts in a standard way
\begin{eqnarray}
\Psi_{[\alpha\beta ]} &=& R_{\alpha\beta} \phi +
\gamma^5_{\alpha\delta} R_{\delta\beta} \widetilde \phi +
\gamma^5_{\alpha\delta} \gamma^\mu_{\delta\tau} R_{\tau\beta} \widetilde
A_\mu \quad,\label{as}\\ \Psi_{\left \{ \alpha\beta \right \}} &=&
\gamma^\mu_{\alpha\delta} R_{\delta\beta} A_\mu
+\sigma^{\mu\nu}_{\alpha\delta} R_{\delta\beta} F_{\mu\nu}\quad,
\label{si}
\end{eqnarray}
where $R= CP$ has the
properties (which are necessary to make expansions (\ref{as},\ref{si}) to
be possible in such a form)
\begin{eqnarray}
&& R^T = -R\quad,\quad R^\dagger =R = R^{-1}\quad,\\
&& R^{-1} \gamma^5 R = (\gamma^5)^T\quad,\\
&& R^{-1} \gamma^\mu R = -(\gamma^\mu)^T\quad,\\
&& R^{-1} \sigma^{\mu\nu} R = - (\sigma^{\mu\nu})^T\quad.
\end{eqnarray}
The  explicit form of this matrix can be chosen:
\begin{equation}
R=\pmatrix{i\Theta & 0\cr 0&-i\Theta\cr}\quad,\quad \Theta = -i\sigma_2 =
\pmatrix{0&-1\cr 1&0\cr},
\end{equation} provided that $\gamma^\mu$
matrices are in the Weyl representation.  The equations
(\ref{bw1},\ref{bw2}) lead to the Kemmer set of the $j=0$ equations:
\begin{eqnarray}
m\phi &=&0 \quad,\\
m\widetilde \phi &=& -i\partial_\mu \widetilde A^\mu\quad,\\
m\widetilde A^\mu &=& -i\partial^\mu \widetilde \phi\quad,
\end{eqnarray}
and to the Proca-Duffin-Kemmer set of the equations for the $j=1$
case:\,\,\footnote{We could use another symmetric matrix $\gamma^5
\sigma^{\mu\nu} R$ in the expansion of the symmetric spinor of the second
rank.  In this case the equations will read
\begin{eqnarray}
&& i\partial_\alpha \widetilde F^{\alpha\mu} +{m\over 2} B^\mu = 0\quad,
\label{de1}\\
&& 2im \widetilde F^{\mu\nu} = \partial^\mu B^\nu -\partial^\nu
B^\mu\quad.\label{de2}
\end{eqnarray}
in which  the dual tensor
$\widetilde F^{\mu\nu}= {1\over 2} \epsilon^{\mu\nu\rho\sigma}
F_{\rho\sigma}$ presents,
because we used that in the Weyl representation
$\gamma^5 \sigma^{\mu\nu} = {i\over 2} \epsilon^{\mu\nu\rho\sigma}
\sigma_{\rho\sigma}$; $B^\mu$ is the corresponding vector potential.  The
equation for the antisymmetric tensor field (which can be obtained from
this set) does not change its form (cf.~[4i,j]) but we see some
``renormalization" of the field functions. In general, it is permitted to
choose various relative phase factors in the expansion of the
symmetric wave function (4) and also consider the matrix term of the form
$\gamma^5 \sigma^{\mu\nu}$.  We shall have additional phase factors in
equations relating the physical fields and the 4-vector potentials.  They
can be absorbed by the redefinition of the potentials/fields.
The above shows that the dual tensor of the
second rank can also be epxanded in potentials. This discussion is
intimately related to the matter of parity properties of $j=1$ field and
with the theoretical quest of existence of {\it antiparticle} in the
$(1,0)\oplus (0,1)$ and/or $(1/2,1/2)$ representations.}
\begin{eqnarray} &&\partial_\alpha
F^{\alpha\mu} + {m\over 2} A^\mu = 0 \quad,\label{1} \\ &&2 m F^{\mu\nu} =
\partial^\mu A^\nu - \partial^\nu A^\mu \quad, \label{2} \end{eqnarray} In
the meantime, in the textbooks, the latter set is usually written as ({\it
e.g.}, p. 135 of ref.~\cite{Itzyk})
\begin{eqnarray}
&&\partial_\alpha F^{\alpha\mu} + m^2 A^\mu = 0 \quad, \label{3}\\ &&
F^{\mu\nu} = \partial^\mu A^\nu - \partial^\nu A^\mu \quad, \label{4}
\end{eqnarray} The set (\ref{3},\ref{4}) is obtained from
(\ref{1},\ref{2}) after the normalization change $A_\mu \rightarrow 2m
A_\mu$ or $F_{\mu\nu} \rightarrow {1\over 2m} F_{\mu\nu}$.  Of course, one
can investigate other sets of equations with different normalization of
the $F_{\mu\nu}$ and $A_\mu$ fields. Are all these sets of equations
equivalent?  As we shall see, to answer this question is not trivial.
Ahluwalia argued that the physical normalization is such that in the
massless-limit zero-momentum field functions should vanish in the momentum
representation (there are no massless particles at rest). We
advocate the following approach:  the massless limit can and must be taken
in the end of all calculations only, {\it i.~e.}, for physical quantities.

Let us proceed further. In order to be able to answer the question about
the behaviour of the spin operator
${\bf J}^i = {1\over 2} \epsilon^{ijk}
J^{jk}$ in the massless limit
one should know the behaviour of the fields $F_{\mu\nu}$ and/or $A_\mu$ in
the massless limit.  We want to analyze the first set (\ref{1},\ref{2}).
If one advocates the following definitions (p. 209 of~\cite{Wein})
\begin{eqnarray}
\epsilon^\mu  ({\bf 0}, +1) = - {1\over \sqrt{2}}
\pmatrix{0\cr 1\cr i \cr 0\cr}\,,\quad
\epsilon^\mu  ({\bf 0}, 0) =
\pmatrix{0\cr 0\cr 0 \cr 1\cr}\,,\quad
\epsilon^\mu  ({\bf 0}, -1) = {1\over \sqrt{2}}
\pmatrix{0\cr 1\cr -i \cr 0\cr}
\end{eqnarray}
and ($\widehat p_i = p_i /\mid {\bf p} \mid$,\, $\gamma
= E_p/m$), p. 68 of ref.~\cite{Wein},
\begin{eqnarray} && \epsilon^\mu ({\bf p}, \sigma) =
L^{\mu}_{\quad\nu} ({\bf p}) \epsilon^\nu ({\bf 0},\sigma)\,, \\
&& L^0_{\quad 0} ({\bf p}) = \gamma\, ,\quad L^i_{\quad 0} ({\bf p}) =
L^0_{\quad i} ({\bf p}) = \widehat p_i \sqrt{\gamma^2 -1}\, ,\\
&&L^i_{\quad k} ({\bf p}) = \delta_{ik} +(\gamma -1) \widehat p_i \widehat
p_k \quad \end{eqnarray}
for the 4-vector potential field,\footnote{Remember that the invariant
integral measure over the Minkowski space for physical particles is $$\int
d^4 p \delta (p^2 -m^2)\equiv \int {d^3  {\bf p} \over 2E_p}\quad,\quad
E_p = \sqrt{{\bf p}^2 +m^2}\quad.$$ Therefore, we use the field operator
as in (\ref{fo}). The coefficient $(2\pi)^3$ can be considered at this
stage as chosen for the convenience.  In ref.~\cite{Wein} the factor
$1/(2E_p)$ was absorbed in creation/annihilation operators and instead of
the field operator (\ref{fo}) the operator was used in which the
$\epsilon^\mu ({\bf p}, \sigma)$ functions for a massive spin-1 particle
were substituted by $u^\mu ({\bf p}, \sigma) = (2E_p)^{-1/2} \epsilon^\mu
({\bf p}, \sigma)$, what may lead to confusions in the definitions of the
massless limit $m\rightarrow 0$ for  classical polarization vectors.}
p. 129 of ref.~\cite{Itzyk}
\begin{equation} A^\mu (x^\mu) =
\sum_{\sigma=0,\pm 1} \int {d^3 {\bf p} \over (2\pi)^3 }
{1\over 2E_p} \left
[\epsilon^\mu ({\bf p}, \sigma) a ({\bf p}, \sigma) e^{-ip\cdot x} +
(\epsilon^\mu ({\bf p}, \sigma))^c b^\dagger ({\bf p}, \sigma) e^{+ip\cdot
x} \right ]\,, \label{fo}
\end{equation}
the normalization of the wave
functions in the momentum representation is thus chosen to the unit,
$\epsilon_\mu^\ast ({\bf p}, \sigma) \epsilon^\mu ({\bf p},\sigma) = -
1$.\footnote{The metric used in this paper $g^{\mu\nu} = \mbox{diag} (1,
-1, -1, -1)$ is different from that of ref.~\cite{Wein}.} \, We observe
that in the massless limit all the defined polarization vectors of the
momentum space do not have good behaviour; the functions describing spin-1
particles tend to infinity.  This is not satisfactory.\footnote{
It is interesting to remind that the authors of ref.~\cite{Itzyk}
(see page 136 therein) tried to inforce the Stueckelberg's Lagrangian in
order to overcome the difficulties related with the $m\rightarrow 0$ limit
(or the Proca theory $\rightarrow$ Quantum Electrodynamics).  The
Stueckelberg's Lagrangian is well known to contain an additional term
which may be put in correspondence to some scalar (longitudinal) field
(cf.  also~\cite{Staru}).} Nevertheless,
after renormalizing the potentials, {\it e.~g.}, $\epsilon^\mu \rightarrow
u^\mu \equiv m \epsilon^\mu$ we come to the field functions in the momentum
representation:
\begin{equation}
u^\mu
({\bf p}, +1)= -{N\over \sqrt{2}m}\pmatrix{p_r\cr m+ {p_1 p_r \over
E_p+m}\cr im +{p_2 p_r \over E_p+m}\cr {p_3 p_r \over
E_p+m}\cr}\,,\quad  u^\mu ({\bf p}, -1)= {N\over
\sqrt{2}m}\pmatrix{p_l\cr m+ {p_1 p_l \over E_p+m}\cr -im +{p_2 p_l \over
E_p+m}\cr {p_3 p_l \over E_p+m}\cr}\label{vp12}
\end{equation}
\begin{equation}
\qquad \qquad \qquad u^\mu ({\bf
p}, 0) = {N\over m}\pmatrix{p_3\cr {p_1 p_3 \over E_p+m}\cr {p_2 p_3
\over E_p+m}\cr m + {p_3^2 \over E_p+m}\cr}\,,  \label{vp3}
\end{equation}
($N=m$ and $p_{r,l} = p_1 \pm ip_2$) which do not
diverge in the massless limit.  Two of the massless functions (with
 $h = \pm 1$) are equal to zero when a particle, described by this
field, is moving along the third axis ($p_1 = p_2 =0$,\, $p_3 \neq 0$).
The third one ($h = 0$) is
\begin{equation} u^\mu (p_3, 0)
\mid_{m\rightarrow 0} = \pmatrix{p_3\cr 0\cr 0\cr {p_3^2 \over E_p}\cr}
\equiv  \pmatrix{E_p \cr 0 \cr 0 \cr E_p\cr}\quad, \end{equation}
\setcounter{footnote}{0}
and at
the rest ($E_p=p_3 \rightarrow 0$) also vanishes. Thus, such a field
operator describes the ``longitudinal photons" what is in the complete
accordance with the Weinberg theorem $B-A= h$
for massless particles (we
use the $D(1/2,1/2)$ representation). Thus, the change of the
normalization can lead to the ``change" of physical content described by
the classical field (at least, comparing with the well-accepted one).
In the quantum case one should somehow fix the form of commutation
relations by some physical principles. They may be fixed by
requirements of the dimensionless of the action (apart from the
requirements of the translational and rotational invariancies).

Furthermore, it is easy to prove that the physical
fields $F^{\mu\nu}$ (defined as in (\ref{1},\ref{2}),
for instance) vanish in the
massless zero-momentum limit under the both definitions of normalization
and field equations. It is straightforward to find ${\bf
B}^{(+)} ({\bf p}, \sigma ) = {i\over 2m} {\bf p} \times {\bf u}
({\bf p}, \sigma)$, \, ${\bf E}^{(+)} ({\bf p}, \sigma) = {i\over 2m}
p_0 {\bf u} ({\bf p}, \sigma) - {i\over 2m} {\bf p} u^0
({\bf p}, \sigma)$ and the corresponding negative-energy strengths
for the field operator (in general, complex-valued)
\begin{equation}
F^{\mu\nu} (x)=\sum_{\sigma=0,\pm 1}
\int \frac{d^3 {\bf p}}{(2\pi)^3 2E_p} \, \left [
F^{\mu\nu}_{(+)} ({\bf p},\sigma)\, a ({\bf p},\sigma)\, e^{-ip x} +
F^{\mu\nu}_{(-)} ({\bf p},\sigma) \,b^\dagger ({\bf p},\sigma)\,
e^{+ip x} \right ]\label{fop}
\end{equation}
Here they are:\footnote{In this paper we assume that $[\epsilon^\mu ({\bf
p},\sigma) ]^c =e^{i\alpha_\sigma} [\epsilon^\mu ({\bf p},\sigma )
]^\ast$, with $\alpha_\sigma$ being arbitrary phase factors at this stage.
Thus, ${\cal C} = I_{4\times 4}$ and $S^C ={\cal K}$.
Taking in mind the problem of indefinite metrics
it would be interesting to investigate other choices of the ${\cal C}$,
the charge conjugation matrix and/or consider a field operator composed
of CP-conjugate states.}
\begin{eqnarray}
{\bf B}^{(+)} ({\bf p},
+1) &=& -{iN\over 2\sqrt{2}m} \pmatrix{-ip_3 \cr p_3 \cr ip_r\cr} =
+ e^{-i\alpha_{-1}} {\bf B}^{(-)} ({\bf p}, -1 ) \quad,\quad   \label{bp}\\
{\bf B}^{(+)} ({\bf
p}, 0) &=& {iN \over 2m} \pmatrix{p_2 \cr -p_1 \cr 0\cr} =
- e^{-i\alpha_0} {\bf B}^{(-)} ({\bf p}, 0) \quad,\quad \label{bn}\\
{\bf B}^{(+)} ({\bf p}, -1)
&=& {iN \over 2\sqrt{2} m} \pmatrix{ip_3 \cr p_3 \cr -ip_l\cr} =
+ e^{-i\alpha_{+1}} {\bf B}^{(-)} ({\bf p}, +1)
\quad,\quad\label{bm}
\end{eqnarray}
and
\begin{eqnarray}
{\bf E}^{(+)} ({\bf p}, +1) &=&  -{iN\over 2\sqrt{2}m} \pmatrix{E_p- {p_1
p_r \over E_p +m}\cr iE_p -{p_2 p_r \over E_p+m}\cr -{p_3 p_r \over
E+m}\cr} = + e^{-i\alpha^\prime_{-1}}
{\bf E}^{(-)} ({\bf p}, -1) \quad,\quad\label{ep}\\
{\bf E}^{(+)} ({\bf p}, 0) &=&  {iN\over 2m} \pmatrix{- {p_1 p_3
\over E_p+m}\cr -{p_2 p_3 \over E_p+m}\cr E_p-{p_3^2 \over
E_p+m}\cr} = - e^{-i\alpha^\prime_0} {\bf E}^{(-)} ({\bf p}, 0)
\quad,\quad\label{en}\\
{\bf E}^{(+)} ({\bf p}, -1) &=&  {iN\over
2\sqrt{2}m} \pmatrix{E_p- {p_1 p_l \over E_p+m}\cr -iE_p -{p_2 p_l \over
E_p+m}\cr -{p_3 p_l \over E_p+m}\cr} = + e^{-i\alpha^\prime_{+1}} {\bf
E}^{(-)} ({\bf p}, +1) \quad,\quad\label{em}
\end{eqnarray}
where we denoted, as previously, a normalization factor appearing in the
definitions of the potentials (and/or in the definitions of the physical
fields through potentials) as $N$.

For the sake of completeness let us
present the fields corresponding to the ``time-like" polarization:
\begin{equation}
u^\mu ({\bf p}, 0_t) = {N \over m} \pmatrix{E_p\cr p_1
\cr p_2\cr p_3\cr}\quad,\quad {\bf B}^{(\pm)} ({\bf p}, 0_t) = {\bf
0}\quad,\quad {\bf E}^{(\pm)} ({\bf p}, 0_t) = {\bf 0}\,\,.
\label{tp}
\end{equation}
The polarization vector $u^\mu ({\bf p}, 0_t)$ has
good behaviour in $m\rightarrow 0$, $N=m$ (and also in the subsequent
limit ${\bf p} \rightarrow {\bf 0}$) and it may correspond to some
field (particle).
As one can see the field operator composed of the states
of longitudinal (e.g., as positive-energy solution) and time-like
(e.g., as negative-energy solution)
polarizations may describe a situation when a particle and an antiparticle
have {\it opposite} intrinsic parities.  Furthermore, in the
case of the normalization of potentials to the mass $N=m$  the physical
fields ${\bf B}$ and ${\bf E}$, which correspond to the ``time-like"
polarization, are equal to zero identically.  The longitudinal fields
(strengths) are equal to zero in this limit only when one chooses the
frame  with $p_3 = \mid {\bf p} \mid$, cf. with the light front
formulation, ref.~\cite{DVALF}.  In the case $N=1$ and (\ref{1},\ref{2})
the fields ${\bf B}^\pm ({\bf p},
0_t)$ and ${\bf E}^\pm ({\bf p}, 0_t)$ would be undefined.

\section{Lagrangian, Energy-Momentum Tensor and Angular Momentum}

We begin with the Lagrangian,
including, in general, mass term:\footnote{The massless limit
($m\rightarrow 0$) of the Lagrangian is connected with the Lagrangians
used in the conformal field theory and in the conformal supergravity by
adding the total derivative:
\begin{equation} {\cal L}_{CFT} = {\cal L} +
{1\over 2}\partial_\mu \left ( F_{\nu\alpha} \partial^\nu F^{\mu\alpha} -
F^{\mu\alpha} \partial^\nu F_{\nu\alpha} \right )\quad.  \end{equation}
The Kalb-Ramond gauge-invariant
form (with
respect to ``gauge" transformations $F_{\mu\nu}  \rightarrow F_{\mu\nu}
+\partial_\nu \Lambda_\mu - \partial_\mu \Lambda_\nu$),
ref.~\cite{Ogievet,Hayashi}, is obtained only if one uses the Fermi
procedure {\it mutatis mutandis} by removing the additional ``phase" field
$\lambda (\partial_\mu F^{\mu\nu})^2$, with the appropriate coefficient
$\lambda$, from the Lagrangian. This has certain analogy with the QED,
where the question, whether the Lagrangian is gauge-invariant or not, is
solved depending on the presence of the term $\lambda (\partial_\mu
A^\mu)^2$. For details see ref.~\cite{Hayashi}.

In general it is possible to introduce various forms of the mass term
and of corresponding normalization of the field. But, the dimensionless
of the action ${\cal S}$ may impose some restrictions; we know that
$F^{\mu\nu}$ in order to be able to describe long-range forces should have
the dimension $[energy]^2$. In order to take this into account one should
divide the Lagrangian (\ref{Lagran}) by $m^2$; calculate
corresponding dynamical invariants, other observable quantities; and only
then study $m\rightarrow 0$ limit.}
\begin{equation} {\cal L} =  {1\over
4} (\partial_\mu F_{\nu\alpha})(\partial^\mu F^{\nu\alpha}) - {1\over 2}
(\partial_\mu F^{\mu\alpha})(\partial^\nu F_{\nu\alpha}) - {1\over 2}
(\partial_\mu F_{\nu\alpha})(\partial^\nu F^{\mu\alpha}) + {1\over 4} m^2
F_{\mu\nu} F^{\mu\nu} \,.  \label{Lagran} \end{equation} The Lagrangian
leads to the equation of motion in the following form (provided that the
appropriate antisymmetrization procedure has been taken into account):
\begin{equation} {1\over 2} ({\,\lower0.9pt\vbox{\hrule
\hbox{\vrule height 0.2 cm \hskip 0.2 cm \vrule height
0.2cm}\hrule}\,}+m^2) F_{\mu\nu} +
(\partial_{\mu}F_{\alpha\nu}^{\quad,\alpha} -
\partial_{\nu}F_{\alpha\mu}^{\quad,\alpha}) = 0 \quad,\label{PE}
\end{equation}
where ${\,\lower0.9pt\vbox{\hrule \hbox{\vrule height 0.2 cm
\hskip 0.2 cm
\vrule height 0.2 cm}\hrule}\,}
=- \partial_{\alpha}\partial^{\alpha}$, cf. with the set of equations
(15,16).  It is this equation for antisymmetric-tensor-field components
that follows from the Proca-Duffin-Kemmer-Bargmann-Wigner
consideration
provided that $m\neq 0$ and in the final expression one takes into account
the Klein-Gordon equation $({\,\lower0.9pt\vbox{\hrule \hbox{\vrule height
0.2 cm \hskip 0.2 cm \vrule height 0.2 cm}\hrule}\,} - m^2) F_{\mu\nu}=
0$.  The latter expresses relativistic dispersion relations $E^2 -{\bf p}^2
=m^2$.

Following the variation procedure
one can obtain that the energy-momentum tensor is expressed:
\begin{eqnarray}
\Theta^{\lambda\beta} &=& {1\over 2} \left [
(\partial^\lambda F_{\mu\alpha}) (\partial^\beta F^{\mu\alpha})
- 2(\partial_\mu F^{\mu\alpha}) (\partial^\beta F^\lambda_{\quad\alpha}) -
\right.\nonumber\\
&-& \left . 2 (\partial^\mu F^{\lambda\alpha}) (\partial^\beta
F_{\mu\alpha})\right ] -{\cal L} g^{\lambda\beta}\, .
\end{eqnarray}
One can also obtain that
for rotations $x^{\mu^\prime} = x^\mu + \omega^{\mu\nu} x_\nu$
the corresponding variation of the wave function is found
from the formula:
\begin{equation}
\delta F^{\alpha\beta} = {1\over 2} \omega^{\kappa\tau}
{\cal T}_{\kappa\tau}^{\alpha\beta,\mu\nu} F_{\mu\nu}\quad.
\end{equation}
The generators of infinitesimal transformations are then defined as
\begin{eqnarray}
\lefteqn{{\cal T}_{\kappa\tau}^{\alpha\beta,\mu\nu} \,=\,
{1\over 2} g^{\alpha\mu} (\delta_\kappa^\beta \,\delta_\tau^\nu \,-\,
\delta_\tau^\beta\,\delta_\kappa^\nu) \,+\,{1\over 2} g^{\beta\mu}
(\delta_\kappa^\nu\delta_\tau^\alpha  \,-\,
\delta_\tau^\nu\, \delta_\kappa^\alpha) +\nonumber}\\
&+&\,
{1\over 2} g^{\alpha\nu} (\delta_\kappa^\mu \, \delta_\tau^\beta \,-\,
\delta_\tau^\mu \,\delta_\kappa^\beta) \,+\, {1\over 2}
g^{\beta\nu} (\delta_\kappa^\alpha \,\delta_\tau^\mu \,-\,
\delta_\tau^\alpha \, \delta_\kappa^\mu)\quad.
\end{eqnarray}
It is ${\cal T}_{\kappa\tau}^{\alpha\beta,\mu\nu}$, the generators of
infinitesimal transformations,
that enter in the formula for the relativistic spin tensor:
\begin{equation}
J_{\kappa\tau} = \int d^3 {\bf x} \left [ \frac{\partial {\cal
L}}{\partial ( \partial F^{\alpha\beta}/\partial t )} {\cal
T}^{\alpha\beta,\mu\nu}_{\kappa\tau} F_{\mu\nu} \right ]\quad.
\label{inv}
\end{equation}
As a result one  obtains:
\begin{eqnarray}
J_{\kappa\tau} &=& \int d^3 {\bf x} \left [ (\partial_\mu F^{\mu\nu})
(g_{0\kappa} F_{\nu\tau} - g_{0\tau} F_{\nu\kappa}) -  (\partial_\mu
F^\mu_{\,\,\,\,\kappa}) F_{0\tau} + (\partial_\mu F^\mu_{\,\,\,\,\tau})
F_{0\kappa} + \right. \nonumber\\
&+& \left. F^\mu_{\,\,\,\,\kappa} ( \partial_0 F_{\tau\mu} +
\partial_\mu F_{0\tau} +\partial_\tau F_{\mu 0})  -   F^\mu_{\,\,\,\,\tau}
( \partial_0 F_{\kappa\mu} +\partial_\mu F_{0\kappa} +\partial_\kappa
F_{\mu 0}) \right ]\,. \label{gene10}
\end{eqnarray}
If one agrees that the
orbital part of the angular momentum
\begin{equation} L_{\kappa\tau} =
x_\kappa \Theta_{0\,\tau} - x_\tau \Theta_{0\,\kappa} \quad,
\end{equation}
with  $\Theta_{\tau\lambda}$ being the energy-momentum tensor, does not
contribute to the Pauli-Lubanski operator when acting on the
one-particle free states (as in the Dirac $j=1/2$ case), then
the Pauli-Lubanski 4-vector is constructed as
follows, Eq. (2-21) of~\cite{Itzyk}
\begin{equation}
W_\mu = -{1\over 2}  \epsilon_{\mu\kappa\tau\nu} J^{\kappa\tau} P^\nu \quad,
\end{equation}
with $J^{\kappa\tau}$ defined by Eqs.
(\ref{inv},\ref{gene10}). The 4-momentum operator $P^\nu$ can be replaced
by its eigenvalue when acting on the plane-wave eigenstates.

Furthermore, one should
choose space-like normalized vector $n^\mu n_\mu = -1$, for example $n_0
=0$,\, ${\bf n} = \widehat  {\bf p} = {\bf p} /\vert {\bf
p}\vert$.\,\,\footnote{One should remember that the helicity operator is
usually connected with the Pauli-Lubanski vector in the following manner
$({\bf J} \cdot \widehat {\bf p}) = ({\bf W} \cdot \widehat {\bf p})/
E_p$, see ref.~\cite{Shirok}. The choice of ref.~\cite{Itzyk}, p. 147,
$n^\mu = \left ( t^\mu - p^\mu {p\cdot t \over m^2} \right ) {m\over \mid
{\bf p} \mid}$, with $t^\mu \equiv (1,0,0,0)$ being a time-like vector, is
also possible but it leads to some oscurities in the procedure of taking
the massless limit. These oscurities will be clarified in a separate
paper.} \,\, After lengthy calculations in a spirit of pp.
58, 147 of~\cite{Itzyk}, one can find the explicit form of
the relativistic spin:
\begin{eqnarray} && (W_\mu \cdot n^\mu) = - ({\bf
W}\cdot {\bf n}) = -{1\over 2} \epsilon^{ijk} n^k J^{ij}
p^0\quad,\label{PL1}\\
&& {\bf J}^k = {1\over 2} \epsilon^{ijk} J^{ij} =
\epsilon^{ijk} \int d^3 {\bf x} \left [ F^{0i} (\partial_\mu F^{\mu j}) +
F_\mu^{\,\,\,\,j} (\partial^0 F^{\mu i} +\partial^\mu F^{i0} +\partial^i
F^{0\mu} ) \right ]\,.\nonumber\\
&&\label{PL2}
\end{eqnarray}
Now it becomes obvious that the application of the generalized Lorentz
conditions (which are quantum versions of free-space dual Maxwell's
equations) leads in such a formulation to the absence of electromagnetism
in a conventional sense.  The resulting Kalb-Ramond field is longitudinal
(helicity $h=0$).  All the components of the angular momentum tensor
for this case are identically equated to zero.

One can consider connections with earlier works
and regard $\sim {\bf A}\times {\bf A}^\ast$ term  as the
part of antisymmetric tensor potential and $\sim {\bf B}\times {\bf
B}^\ast$, as the part of the 4-vector field.
According to~\cite[Eqs.(9,10)]{Ogievet} we proceed in the
construction of the ``potentials" for the notoph
(Ogievetski\u{\i}-Polubarino-Kalb-Ramond field) as follows:
\begin{equation}
\tilde F_{\mu\nu} ({\bf p}) = N \left [\epsilon_\mu^{(1)} ({\bf
p})\epsilon_\nu^{(2)} ({\bf p})- \epsilon_\nu^{(1)} ({\bf p})
\epsilon_\mu^{(2)} ({\bf p}) \right ] \end{equation}
On using explicit
forms for the polarization vectors in the momentum space  one obtains
\begin{eqnarray}
\tilde F^{\mu\nu} ({\bf p}) = {iN^2 \over m} \pmatrix{0&-p_2&
p_1& 0\cr p_2 &0& m+{p_r p_l\over p_0+m} & {p_2 p_3\over p_0 +m}\cr -p_1 &
-m - {p_r p_l \over p_0+m}& 0& -{p_1 p_3\over p_0 +m}\cr 0& -{p_2 p_3
\over p_0 +m} & {p_1 p_3 \over p_0+m}&0\cr}\, , \label{lc} \end{eqnarray}
i.e., it coincides with the longitudinal components of the antisymmetric
tensor obtained in refs.~\cite[Eqs.(2.14,2.17)]{Ahlu}
and my previous works within the normalization and
different forms of the spin basis (cf. formulas (29,32) above).  The
longitudinal states reduce to zero in the massless case under appropriate
choice of the normalization and only if a $j=1$ particle moves along with
the third axis $OZ$. It is also useful to compare Eq. (\ref{lc}) with the
formula (B2) in ref.~\cite{DVALF} in order to realize the correct
procedure for taking the massless limit.

Finally, we agree with the previous authors, e.~g., ref.~\cite{Ohanian},
see Eq. (4) therein, about the gauge {\it non}-invariance of
the division of the angular momentum of the electromagnetic field into the
``orbital" and ``spin" part (\ref{PL2}). We proved again
that for the antisymmetric tensor field ${\bf J} \sim \int d^3 {\bf x}\,
{\bf E}\times {\bf A}$. So, what people actually did (when spoken about
the Ogievetski\u{\i}-Polubarinov-Kalb-Ramond field) is:
When $N=m$ they considered the gauge part of the 4-vector field functions.
Then, equated ${\bf A}$ containing the transverse modes on choosing
$p_r =p_l =0$ in the massless limit
(see formulas (\ref{vp12})).\footnote{The reader, of course,
can consider equating by the usual gauge transformation, $A^\mu
\rightarrow A^\mu +\partial^\mu \chi$.} Under this choice the ${\bf E}
({\bf p}, 0)$ and ${\bf B} ({\bf p}, 0)$ are equal to zero in massless
limit.  But, the gauge part of $u^\mu ({\bf p}, 0)$ is not. The spin
angular momentum can still be zero.
When $N=1$ the situation is the same because of the different form
of dynamical equations. So, for those who prefer simpler consideration it
is enough to regard all possible states of 4-potentials/antisymmetric
tensor field in the massless limit in the calculation of physical
observables. Of course, I would like to repeat, it is not yet clear
and it is not yet supported by reliable experiments whether
the third state of the 4-vector potential/antisymmetric tensor field has
{\it physical} significance and whether it is observable.

However, in~\cite{poten} and several previous works experiments for
verification of {\it real physical} significance of 4-potentials have been
designed.

\section{Conclusions}

The achieved conclusion is: both the  antisymmetric tensor field and
the 4-potential field may have third helicity state in massless limit
from a theoretical viewpoint.
This problem is connected with the problem of the observability of
the gauge.  This conclusion is achieved on the basis of the analysis of
the problem from a viewpoint of the normalization.

\section*{Acknowledgments}

This work was partially supported by the Mexican Sistema Nacional de
Investigadores and the Programa de Apoyo a la Carrera Docente.

I thank the sponsors of the Workshop (SEP, NSF-CONACyT, CLAF, CLAF-M),
whose help made it possible.

I greatly appreciate useful information from
referees of {\it Helvetica Physica Acta}, {\it Foundation of Physics},
{\it Physical Review}, and from Profs.  A. Chubykalo, V. Dubovik, S.
Esposito, A. Nikitin, A.  Pashkov, E. Recami. Frank discussions with D.
Ahluwalia (1993-1998) are acknowledged, even if I do {\it not} always
agree with him.

\end{document}